\documentclass[aap,seceqn,citesort,MSNbibl,dvips]{arximspdf}
\usepackage{mathrsfs}

\doi{10.1214/09-AAP648}
\volume{20}
\issue{4}
\pubyear{2010}
\firstpage{1341}
\lastpage{1358}

\makeatletter

\newproclaim{defi}{Definition}[section]
\newproclaim{bem}[defi]{Remark}
\newtheorem{lemma}[defi]{Lemma}
\newtheorem{prop}[defi]{Proposition}
\newtheorem{theorem}[defi]{Theorem}

\newproclaim{Step}{Step}

\newcommand{\auf}{[\![}
\newcommand{\zu}{]\!]}

\makeatother

\begin{document}
\begin{frontmatter}

\title{On using shadow prices in portfolio optimization with
transaction costs}
\runtitle{Portfolio optimization with transaction costs}

\begin{aug}
\author[A]{\fnms{J.} \snm{Kallsen}\corref{}\ead[label=e1]{kallsen@math.uni-kiel.de}} and
\author[B]{\fnms{J.} \snm{Muhle-Karbe}\ead[label=e2]{johannes.muhlekarbe@univie.ac.at}}
\runauthor{J. Kallsen and J. Muhle-Karbe}
\affiliation{Christian-Albrechts-Universit\"{a}t zu Kiel and
Universit\"
{a}t Wien}
\address[A]{Mathematisches Seminar\\
Christian-Albrechts-Universit\"{a}t zu Kiel\\
Christian-Albrechts-Platz 4\\
D-24098 Kiel\\
Germany\\
\printead{e1}}
\address[B]{Fakult\"{a}t f\"{u}r Mathematik\\
Universit\"{a}t Wien\\
Nordbergstrasse 15\\
A-1090 Wien\\
Austria \\
\printead{e2}}
\end{aug}

\received{\smonth{4} \syear{2008}}
\revised{\smonth{5} \syear{2009}}

%
\begin{abstract}
In frictionless markets, utility maximization problems are typically
solved either by stochastic control or by martingale methods. Beginning
with the seminal paper of Davis and Norman [\textit{Math. Oper. Res.}
\textbf{15} (1990) 676--713],
stochastic control theory has also been used to solve various problems
of this type in the presence of proportional transaction costs.
Martingale methods, on the other hand, have so far only been used to
derive general structural results. These apply the duality theory for
frictionless markets typically to a fictitious \textit{shadow price
process} lying within the bid-ask bounds of the real price process.

In this paper, we show that this dual approach can actually be used for
both deriving a candidate solution and verification in Merton's problem
with logarithmic utility and proportional transaction costs. In
particular, we determine the shadow price process.
\end{abstract}

%

\end{frontmatter}

\section{Introduction}\label{intro}
A basic question in mathematical finance is how to choose an optimal
investment strategy in a securities market or, more specifically, how
to maximize utility from consumption.
This is often called the \textit{Merton problem} because it was solved
by Merton \cite{merton69,merton71} for power and logarithmic utility
functions in a Markovian It\^{o} process model. In a market with a
riskless bank account and one risky asset following a geometric
Brownian motion, the optimal strategy turns out to invest a
\textit{constant fraction} $\pi{}^*$ of wealth in the risky asset and to consume
at a rate proportional to current wealth. This means that it is optimal
for the investor to keep her portfolio holdings in bank and stock on
the so-called \textit{Merton line} with slope $\pi^*/(1-\pi^*)$.

In a continuous time setting, proportional transaction costs were
introduced to the Merton problem by Magill and Constantinides \cite
{magillconstantinides76}. Their paper contains the fundamental
insight that it is optimal to refrain from transacting while the
portfolio holdings remain in a wedge-shaped \textit{no-transaction
region}, that is, while the fraction of wealth held in stock lies inside
some interval $[\pi_1^*,\pi_2^*]$. However, their solution is derived
in a somewhat heuristic way and also did not show how to compute the
location of the boundaries $\pi_1^*$, $\pi_2^*$.

Mathematically rigorous results were first obtained in the seminal
paper of Davis and Norman \cite{davisnorman90}. They show that it is
indeed optimal to keep the proportion of total wealth held in stock
between fractions $\pi_1^*$, $\pi_2^*$ and they also prove that these
two numbers can be determined as the solution to a free boundary value
problem. The theory of viscosity solutions to Hamilton--Jacobi--Bellman
equations was introduced to this problem by Shreve and Soner \cite
{shrevesoner94} who succeeded in removing several assumptions needed
in \cite{davisnorman90}.

These articles aiming for the computation of the optimal portfolio
employ tools from stochastic control. It seems that unlike for
frictionless markets, martingale methods have so far only been used to
obtain structural existence results in the presence of transaction
costs. In this context, the martingale and duality theory for
frictionless markets is often applied to a \textit{shadow price process}
$\widetilde{S}$ lying within the bid-ask bounds of the real price
process $S$.
Economically speaking, the frictionless price process $\widetilde{S}$
and the
original price process $S$ with transaction costs
lead to identical decisions and gains for the investor under consideration.
This concept has been used in the context of the Fundamental theorem of
Asset Pricing (cf. \cite{jouinikallal95} and recently \cite
{guasonial07,guasonial07b}), local risk minimization \cite
{lambertonal98}, super-replication \cite
{cvitanical99c,bouchardtouzi00} and utility maximization \cite
{cvitanickaratzas96b,cvitanicwang01,loewenstein00}.

In the present study, we reconsider Merton's problem for logarithmic
utility and under proportional transaction costs as in
\cite{davisnorman90}. Our goal is threefold. Most importantly, we show
that the shadow price approach can be used to come up with a candidate
solution to the utility maximization problem under transaction costs.
Moreover, the ensuing verification procedure appears---at least for the
problem at hand---to be relatively simple compared to the very
impressive and nontrivial reasoning in \cite {davisnorman90} and
\cite{shrevesoner94}. Finally, we also construct the shadow price as
part of the solution. For a recent application of the approach of the
current paper, we refer to~\cite{kuehnstroh09}.

The more involved case of power utility is treated in \cite
{davisnorman90,shrevesoner94} as well. The application of the
present approach to this case is subject of current research. While it
is still possible to come up with a candidate for the shadow price, the
corresponding free boundary problem appears to be more difficult than
its counterpart in \cite{davisnorman90}. This stems from the fact
that it may be more difficult to determine the shadow price than the
optimal strategy for power utility (cf. Remark \ref{b:connectDN} for
more details).

The remainder of the paper is organized as follows. The setup is
introduced in Section \ref{s:setup}. Subsequently, we heuristically
derive the free-boundary problem that characterizes the solution.
Verification is done in Section \ref{s:proofs}.

\section{The Merton problem with transaction costs}\label{s:setup}

We study the problem of maximizing expected logarithmic utility from
consumption over an infinite horizon in the presence of proportional
transaction costs. Except for a slightly larger class of admissible
strategies, we work in the setup of \cite{davisnorman90}.

The mathematical framework is as follows: fix a complete, filtered
probability space $(\Omega,\mathscr F,(\mathscr F_t)_{t \in\mathbb R
_+},P)$ supporting a
standard Brownian motion $(W_t)_{t \in\mathbb R _+}$. Our market
consists of
two investment opportunities: a bank account or bond with constant
value $1$ and a risky asset (``stock'') whose discounted price process
$S$ is modelled as a geometric Brownian motion, that is,
%
%
\begin{equation}\label{e:price}
S_t:= S_0 \mathscr E(\mu I+\sigma W)_t=S_0\exp  \biggl( \biggl(\mu
-\frac
{\sigma
^2}{2} \biggr) t+\sigma W_t \biggr)
\end{equation}
with $I_t:=t$ and constants $S_0,\sigma>0$, $\mu\in\mathbb R$.
We consider an investor who disposes of an \textit{initial endowment}
$(\eta_B,\eta_S)\in\mathbb R _+^2$, referring to the number of bonds and
stocks, respectively.
Whenever stock is purchased or sold, transaction costs are imposed
equal to a constant fraction of the amount transacted, the fractions
being $\overline\lambda\in[0,\infty)$ on purchase and $\underline
\lambda\in[0,1)$ on sale, not both being equal to zero. Since
transactions of infinite variation lead to instantaneous ruin, we limit
ourselves to the following set of strategies.
\begin{defi}
A \textit{trading strategy} is an $\mathbb R^2$-valued
predictable process $\varphi=(\varphi^0,\varphi^1)$ of finite
variation, where $\varphi_t^0$ and $\varphi_t^1$ denote the number of
shares held in the bank account and in stock at time $t$, respectively.
A (discounted) \textit{consumption rate} is an $\mathbb R _+$-valued, adapted
stochastic process $c$ satisfying $\int_0^t c_s \,ds<\infty$ a.s. for all
$t \geq0$. A pair $(\varphi,c)$ of a trading strategy $\varphi$ and a
consumption rate $c$ is called \textit{portfolio/consumption pair}.
\end{defi}

To capture the notion of a self-financing strategy, we use the
intuition that no funds are added or withdrawn. To this end, we write
the second component $\varphi^1$ of any strategy $\varphi$ as
difference $\varphi^1=\varphi^{\uparrow}-\varphi^{\downarrow}$ of two
increasing processes $\varphi^{\uparrow}$ and $\varphi^{\downarrow}$
which do not grow at the same time.
Moreover, we denote by
%
%
\begin{equation}\label{e:bidask}
\underline{S}:=(1-\underline\lambda)S,\qquad \overline
{S}:=(1+\overline
\lambda)S,
\end{equation}
the bid and ask price of the stock, respectively. The proceeds of
selling stock must be added to the bank account while the expenses from
consumption and the purchase of stock have to be deducted from the bank
account in any infinitesimal period $(t - dt, t]$, that is, we require
%
%
\begin{equation}\label{e:selff1}
d\varphi^0_t=\underline S_{t-}\,d\varphi^{\downarrow}_t - \overline
S_{t-}\,d\varphi^{\uparrow}_t-c_t\,dt
\end{equation}
for \textit{self-financing strategies}. Written in integral terms, this
amounts to the \textit{self-financing condition}
%
%
\begin{equation}\label{e:selff2}
\varphi^0=\varphi^0_0+\int_0^\cdot\underline S_{t-} \,d\varphi
^{\downarrow
}_t - \int_0^\cdot\overline S_{t-} \,d\varphi^{\uparrow}_t-\int
_0^\cdot c_t\,dt.
\end{equation}
In our setup (\ref{e:price}, \ref{e:bidask}), we obviously have
$\underline S_-=\underline S$ and $\overline S_-=\overline S$ but the
above definition makes sense for
discontinuous bid and ask price processes $\underline{S},\overline{S}$
as well.
The second and third term on the right-hand side represent the
cumulative amount of wealth gained selling, respectively, spent buying
stock, while the last term represents cumulated consumption.
\begin{bem}
Partial integration similarly as in \cite{js87}, I.4.49b, shows that
for $\underline{S}=\overline{S}$, we recover the usual self-financing
condition in frictionless markets (cf. \cite{muhlekarbe09} for more details).
\end{bem}

The value of a portfolio is not obvious either because securities have
no unique price. As is common in the literature, we use the value that
would be obtained if the portfolio were to be liquidated immediately.
\begin{defi}\label{d:optimal}
The (\textit{liquidation}) \textit{value process} of a trading strategy $\varphi$
is defined as
\[
V(\varphi):=\varphi^0+(\varphi^1)^+\underline S -(\varphi
^1)^-\overline S.
\]
A self-financing portfolio/consump\-tion pair $(\varphi,c)$ is called
\textit{admissible} if
$(\varphi^0_0$,\break$\varphi^1_0)=(\eta_B,\eta_S)$ and $V(\varphi) \geq0$.
An admissible pair $(\varphi,c)$ is called \textit{optimal} if it maximizes
%
%
\begin{equation}\label{e:DiscountedUtility}
\kappa\mapsto E \biggl(\int_0^{\infty} e^{-\delta t} \log(\kappa
_t)\,dt \biggr)
\end{equation}
over all admissible portfolio/consumption pairs $(\psi,\kappa)$,
where $\delta>0$ denotes a fixed given \textit{impatience rate}.
\end{defi}

Note that the ``true'' price process $S$ is irrelevant for the problem
as it does not appear in the definitions; only the bid and ask prices
$\underline S,\overline S$ matter. Moreover, since $\delta>0$, the
value function of the Merton problem \textit{without} transaction costs
is finite by \cite{davisnorman90}, Theorem 2.1. Hence, it follows
that this holds in the present setup with transaction costs as well.

Our notion of admissible strategies is slightly more general than that
in \cite{davisnorman90,shrevesoner94}. However, it will turn out
later on that the optimal strategies in both sets coincide.
\begin{lemma}
For any admissible policy $(c,L,U)$ in the sense of \cite
{davisnorman90}, there exists a trading strategy $\varphi=(\varphi
^0,\varphi^1)$ such that $(\varphi,c)$ is an admissible
portfolio/con\-
sump\-tion pair.
\end{lemma}
\begin{pf}
The initial endowment in \cite{davisnorman90} can be expressed in
terms of wealth as $(x,y)=(\eta_B,\eta_SS_0)$.
Define $s_0$, $s_1$ as in \cite{davisnorman90}, (3.1), and
set $\varphi^0_t:=s_0(t-)$, $\varphi^1_t:= s_1(t-)/S_t.$ A simple
calculation shows that $((\varphi^0,\varphi^1),c)$ is an admissible
portfolio/con\-sump\-tion pair.
\end{pf}

\section{Heuristic derivation of the solution}\label{s:heuristik}

As indicated in the \hyperref[intro]{Introduction}, the martingale approach relies
decisively on shadow price processes, which we define as follows.
\begin{defi}
We call a semimartingale $\widetilde S$ \textit{shadow price process} if
%
%
\begin{equation} \label{e:dazwischen}
\underline S \leq\widetilde{S} \leq\overline S
\end{equation}
and if the maximal expected utilities for $S,\underline\lambda
,\overline
\lambda$ and for the price process $\widetilde S$ \textit{without} transaction
costs coincide.
\end{defi}

Obviously, the maximal expected utility for any frictionless price
process $\widetilde S$ satisfying (\ref{e:dazwischen})
is at least as high as for the original market with transaction costs,
since the investor is always
buying at $\widetilde{S}_t \leq\overline S_t$ and selling for
$\widetilde{S}_t
\geq\underline S_t$.

A shadow price process can be interpreted as a kind of least favourable
frictionless market extension.
The corresponding optimal portfolio trades only when the shadow price
happens to coincide with the
bid or ask price, respectively. Otherwise, it would achieve higher
profits with $\widetilde S$ than with
$S$ and transaction costs.

Let us assume that such a shadow price process $\widetilde S$ exists.
If it
were known in the first place,
it would be of great help because portfolio selection problems without
transaction
costs are considerably easier to solve. But it is not known at this stage.
Hence, we must solve the problems of determining $\widetilde S$ and of
portfolio optimization
relative to $\widetilde S$ simultaneously.

To this end, we parametrize the shadow price process in the following form:
%
%
\begin{equation}\label{e:shadowprice}
\widetilde{S}=S\exp(C)
\end{equation}
with some $[\underline{C},\overline{C}]$-valued process $C$ where
\[
\underline{C}:=\log(1-\underline\lambda) \quad\mbox{and}\quad
\overline
{C}:=\log(1+\overline\lambda).
\]
Since $S$ is an It\^{o} process, we expect $\widetilde S$ and hence $C$
to be
It\^{o} processes as well.
We even guess that $C$ is an It\^{o} diffusion, that is,
%
%
\begin{equation}\label{e:defC}
dC_t=\widetilde{\mu}(C_t)\,dt+\widetilde{\sigma}(C_t)\,dW_t
\end{equation}
with some deterministic functions $\widetilde{\mu},\widetilde{\sigma}$.
Any admissible portfolio/consumption pair $(\varphi,c)$ is completely
determined by
$c$ and the fraction of wealth invested in stocks
%
%
\begin{equation}\label{e:pi}
\widetilde\pi:=\frac{\varphi^1 \widetilde S}{\varphi^0+\varphi
^1\widetilde S},
\end{equation}
where bookkeeping is done here relative to shadow prices $\widetilde S$.
Hence, we must determine four unknown objects, namely the ansatz
functions $\widetilde{\mu}$, $\widetilde{\sigma}$ as well as the optimal
consumption rate $c$ and the optimal fraction $\widetilde\pi$ of
wealth in stocks.

Standard results yield the optimal strategy for the frictionless price
process $\widetilde S$.
For example, by \cite{kallsengoll99a}, Theorem 3.1, we have
%
%
\begin{equation}\label{e:optstrat}
\widetilde\pi=\frac{\mu-{\sigma^2}/{2}+\widetilde{\mu
}(C)}{(\sigma
+\widetilde{\sigma
}(C))^2}+\frac{1}{2},
\qquad  c=\delta\widetilde{V}(\varphi),
\end{equation}
where
%
%
\begin{equation}\label{e:schattenwealth}
\widetilde{V}(\varphi)=\varphi^0+\varphi^1\widetilde S
\end{equation}
denotes the value process of $\varphi$ in the frictionless market with
price process $\widetilde{S}$.
This already determines the optimal consumption rate. To simplify the
following calculations, we assume $\widetilde\pi>0$ and work with
\[
\beta:=\log  \biggl(\frac{\widetilde\pi}{1-\widetilde\pi} \biggr)
\]
instead of $\widetilde\pi$. By (\ref{e:pi}) this implies $\beta
:=\log
(\varphi
^1)+\log(\widetilde S)-\log(\varphi^0)$.

Since the optimal strategy trades the shadow price process only when it
coincides with bid or ask price,
$\varphi^1$ must be constant on $\zu0,T\auf$ with $T:=\inf \{
t> 0\dvtx
C_t\in\{\underline{C},\overline{C}\} \}$. By (\ref{e:selff2}) and
It\^{o}'s formula, we have
\[
d\log(\varphi^0_t)=\frac{-c_t}{\varphi^0_t}\,dt
=\frac{-\delta\widetilde{V}_t(\varphi)}{\widetilde{V}_t(\varphi
)-\widetilde\pi_t
\widetilde
{V}_t(\varphi)}\,dt=\frac{-\delta}{1-\widetilde\pi_t}\,dt
\]
on $\zu0,T\auf$, hence insertion of (\ref{e:optstrat}) yields
%
%
\begin{eqnarray}\label{e:Beta1}\hspace*{28pt}
d\beta_t &=& d\log(\varphi^1_t)+d\log(\widetilde S_t)-d\log
(\varphi
^0_t)\nonumber\\
&=& \biggl(\mu-\frac{\sigma^2}{2}+\widetilde{\mu}(C_t)+\frac
{\delta
(\sigma+\widetilde
{\sigma}(C_t))^2}{{1}/{2}(\sigma+\widetilde{\sigma
}(C_t))^2-(\mu
-{\sigma^2}/{2}+\widetilde{\mu}(C_t))} \biggr)\,dt\\
&&{}  +\bigl(\sigma+\widetilde{\sigma}(C_t)\bigr)\,dW_t.\nonumber
\end{eqnarray}
On the other hand, we know from (\ref{e:optstrat}) that $\widetilde
\pi$ is a
function of $C$, which in turn yields $\beta=f(C)$ for some function
$f$. By It\^{o}'s formula, this implies
%
%
\begin{equation}\label{e:Beta2}
d\beta_t= \biggl(f'(C_t)\widetilde{\mu}(C_t)+f''(C_t)\frac
{\widetilde{\sigma
}(C_t)^2}{2} \biggr)\,dt+f'(C_t)\widetilde{\sigma}(C_t)\,dW_t.
\end{equation}
From (\ref{e:Beta1}), (\ref{e:Beta2}) and (\ref{e:optstrat}), we now
obtain three conditions for the three functions $\widetilde{\mu},
\widetilde
{\sigma},f$:
%
%
\begin{eqnarray}
\label{e:cond1}
&&\hspace*{-19pt}\frac{1}{1+e^{-f}}=\frac{\mu-{\sigma^2}/{2}+\widetilde{\mu
}}{(\sigma+\widetilde
{\sigma})^2}+\frac{1}{2},\\
\label{e:cond2}\hspace*{37pt}
&&\mu-\frac{\sigma^2}{2}+\widetilde{\mu}+\frac{\delta(\sigma
+\widetilde
{\sigma
})^2}{{1}/{2}(\sigma+\widetilde{\sigma})^2-(\mu-{\sigma
^2}/{2}+\widetilde
{\mu})}\nonumber\\[-8pt]\\[-8pt]
&&\qquad=f'\widetilde{\mu}+f''\frac{\widetilde{\sigma
}^2}{2},\nonumber\\
\label{e:cond3}
&&\hspace*{-7.65pt}\sigma+\widetilde{\sigma}=f'\widetilde{\sigma}.
\end{eqnarray}
Equations (\ref{e:cond3}) and (\ref{e:cond1}) yield
%
%
\begin{equation}\label{e:sol}
\widetilde{\sigma}=\frac{\sigma}{f'-1},\qquad  \widetilde{\mu
}=- \biggl(\mu
-\frac
{\sigma^2}{2} \biggr)+\frac{\sigma^2}{2} \biggl(\frac
{f'}{f'-1}
\biggr)^2\frac{1-e^{-f}}{1+e^{-f}}.
\end{equation}
By inserting into (\ref{e:cond2}), we obtain the following ordinary
differential equation (ODE) for $f$:
%
%
\begin{eqnarray}\label{e:odef}\quad
f''(x) &=& \frac{2 \delta}{\sigma^2}\bigl(1+e^{f(x)}\bigr)+ \biggl(\frac{2 \mu
}{\sigma^2}-1-\frac{4 \delta}{\sigma^2}\bigl(1+e^{f(x)}\bigr) \biggr)f'(x)
\nonumber\\
& &{} + \biggl(-\frac{4 \mu}{\sigma^2}+\frac{2}{1+e^{-f(x)}}+1+\frac{2
\delta}{\sigma^2}\bigl(1+e^{f(x)}\bigr) \biggr)(f'(x))^2\\
& &{} + \biggl(\frac{2 \mu}{\sigma^2}-\frac{2}{1+e^{-f(x)}}
\biggr)(f'(x))^3.\nonumber
\end{eqnarray}

Because of missing boundary conditions, (\ref{e:odef}) does not yet
yield the solution.
We obtain such conditions heuristically as follows. In order to lead to
finite maximal expected utility, the shadow price process should be
arbitrage-free and hence allow for an equivalent martingale measure.
This in turn means that $\widetilde S$ and hence also $C$ should not
have any
singular part in their semimartingale decomposition.
Put differently, we expect the It\^{o} process representation (\ref
{e:defC}) to hold even
when $C$ reaches the boundary points $\underline{C},\overline{C}$.

The number of shares of stock $\varphi^1$, on the other hand,
changes only when $C$ hits the boundary. As this is likely to happen
only on
a Lebesgue-null set of times, $\varphi^1$~must have a singular part
in order to move at all.
In view of the connection between $\varphi^1$ and $\beta$, this
suggests that $\beta$ has a
singular part as well.
This means that $f$ cannot be a $C^2$ function on the closed interval
$[\underline{C},\overline{C}]$
because otherwise $\beta=f(C)$ would be an It\^{o} process, too.
A natural way out is the ansatz $f'(\underline{C})=-\infty
=f'(\overline
{C})$ in order for
$\beta$ to have a singular part at the boundary.
Hence, we complement ODE (\ref{e:odef})
by boundary conditions
%
%
\begin{equation}\label{e:rand}
\lim_{x\downarrow\underline{C}}f'(x)=-\infty=\lim_{x\uparrow
\overline
{C}}f'(x).
\end{equation}
In order to avoid infinite derivatives, we consider instead the inverse
function \mbox{$g:=f^{-1}$}.
Equation (\ref{e:odef}) turns into
%
%
\begin{eqnarray}\label{e:odeg}\hspace*{28pt}
g''(y) &=& \biggl(-\frac{2 \mu}{\sigma^2}+\frac{2}{1+e^{-y}}
\biggr)\nonumber\\
&&{} + \biggl(\frac{4\mu}{\sigma^2}-\frac{2}{1+e^{-y}}-1-\frac{2 \delta
}{\sigma
^2}(1+e^y) \biggr)g'(y)\\
&&{} + \biggl(-\frac{2\mu}{\sigma^2}+1+\frac{4 \delta}{\sigma
^2}(1+e^y)\biggr)(g'(y))^2 -\frac{2
\delta}{\sigma^2}(1+e^y)(g'(y))^3\nonumber
\end{eqnarray}
on the a priori unknown interval
$[\underline{\beta},\overline{\beta}]:=[f(\overline
{C}),f(\underline{C})]$
and (\ref{e:rand}) translates into free boundary conditions
%
%
\begin{equation}\label{e:randg}
g(\underline{\beta})=\overline{C},\qquad   g(\overline{\beta
})=\underline
{C},\qquad
g'(\underline{\beta})=0, \qquad g'(\overline{\beta})=0.
\end{equation}
Equations (\ref{e:odeg}), (\ref{e:randg}) together with
(\ref{e:shadowprice})--(\ref{e:schattenwealth}) and $f=g^{-1}$
constitute our ansatz for the portfolio optimization problem.

In summary, the solution to the free boundary problem (\ref{e:odeg}),
(\ref{e:randg})---or equivalently (\ref{e:odef}), (\ref{e:rand})---leads
to the optimal strategy. The ODE itself is derived based on the
optimality of $\widetilde\pi$ for $\widetilde S$ and the constancy of
$\varphi^1$
on $\zu0,T\auf$.
In the next section, we show that this ansatz indeed yields the true solution.

Our result resembles \cite{davisnorman90} in that the solution is
expressed in terms of a free boundary problem.
However, both the ODE and the boundary conditions are different, since
the function $g$ refers to the shadow price process from the present
dual approach and therefore does not appear explicitly in the framework
of \cite{davisnorman90} (but cf. Remark \ref{b:connectDN}).

\section{Construction of the shadow price process}\label{s:proofs}

We turn now to verification of the candidate solution from the previous section.
The idea is rather simple. Using (\ref{e:shadowprice}), (\ref{e:defC}),
we define a candidate shadow price process $\widetilde S$.
In order to prove that it is indeed a shadow price process, we show
that the optimal portfolio relative to $\widetilde S$ trades
only at the boundaries $\underline S,\overline S$ .
However, existence of a solution to stochastic differential equation
(SDE) (\ref{e:defC}) is not immediately obvious.
Therefore, we consider instead the corresponding Skorokhod SDE for
$\beta=f(C)$ with instantaneous reflection at some boundaries
$\underline{\beta}<\overline{\beta}$.
The process $C=g(\beta)$ is then defined in a second step.

We begin with an existence result for the free boundary value problem
derived above. We make the following assumption which guarantees that
the fraction of wealth held in stock remains positive and which is
needed in \cite{davisnorman90} as well [(5.1) in that paper].

\subsection*{Standing assumption}
%
%
\begin{equation} \label{e:standingassumption}
0<\mu<\sigma^2.
\end{equation}
\begin{bem}
It is shown in \cite{shrevesoner94} that this condition is not needed
to ensure the existence of an optimal strategy characterized by a
wedge-shaped no-transaction region. 
If the transformation $\beta=\log(\widetilde\pi/(1-\widetilde\pi
))$ was not
used in
our approach, we would still obtain a free boundary problem, but as in
\cite{davisnorman90} it is less obvious whether or not it admits a solution.
\end{bem}
\begin{prop}\label{p:FBVP}
There exist $\underline{\beta} < \overline{\beta}$ and a strictly
decreasing mapping $g\dvtx [\underline{\beta},\overline{\beta}] \to
[\underline{C},\overline{C}]$
satisfying the free boundary problem (\ref{e:odeg}), (\ref{e:randg}).
\end{prop}
\begin{pf}
Since we\vspace*{-3pt} have assumed $0<\frac{\mu}{\sigma^2}<1$, there is a unique
solution $y$ to $\frac{2}{1+e^{-y}}-\frac{2\mu}{\sigma^2}=0$,
namely\vspace*{1pt}
$y_0=-\log({\sigma^2\over\mu}-1)$. For any $\underline\beta
_{\Delta
}:=y_0-\Delta$ with $\Delta>0$, there exists a local solution
$g_{\Delta
}$ of the initial value problem corresponding to (\ref{e:odeg}) and
initial values $g_{\Delta}(\underline\beta_\Delta)=\overline{C}$ and
$g_{\Delta}'(\underline\beta_\Delta)=0$. Set
\[
M':=\max \Biggl\{ \sqrt[3\,]{\frac{4(\mu+\sigma^2)}{\delta}}, \sqrt
{\frac
{8\mu}{\delta}},8+\frac{4\mu+2\sigma^2}{\delta}  \Biggr\}.
\]
Then we have $g_{\Delta}''(y)>0$ for $g_{\Delta}'(y)<-M'$ and
$g_{\Delta
}''(y)<0$ for $g'_{\Delta}>M'$ by (\ref{e:odeg}). Therefore,
$g'_{\Delta
}$ only takes values in $[-M',M']$, which implies that $g_{\Delta}$
does not explode.

From (\ref{e:odeg}) and $\Delta>0$, it follows that $g_{\Delta}''(y)<0$
in a neighborhood $\mathscr{U}$ of $\underline\beta_\Delta$ and hence
$g_{\Delta}'(y)<0$ in $\mathscr{U}$. For sufficiently large $y$ and
$g_{\Delta}'(y)<0$, the right-hand side of (\ref{e:odeg}) is positive
and bounded away from zero by a positive constant. Hence, a comparison
argument shows that there exist further zeros of $g_{\Delta}'$, the
first of which we denote by $\overline{\beta}_{\Delta}$. Note that by
definition $g_{\Delta}$ is strictly decreasing on $[\underline\beta
_\Delta,\overline{\beta}_{\Delta}]$. It remains to show that for
properly chosen $\Delta$, we can achieve $g(\overline{\beta}_{\Delta
})=\underline{C}$ for any $\underline{C} < \overline{C}$.
\begin{Step}\label{Step1}
We first show $g_{\Delta}(\overline{\beta}_{\Delta})\to\overline{C}$
as $\Delta\to0$. This can be seen as follows. Observe that for
$|y-y_0|<1$, (\ref{e:odeg}) and $g'_{\Delta}(y) \in[-M',M']$ yield
\begin{eqnarray*}
|g_\Delta''(y)|< M''
&:=&\frac{2\mu}{\sigma^2}+2+ \biggl(\frac{4\mu}{\sigma^2}+3+\frac
{2\delta
}{\sigma^2} (1+e^{y_0+1} ) \biggr)M'\\
&&{} + \biggl(\frac{2\mu}{\sigma^2}+1+\frac{4\delta}{\sigma^2}
(1+e^{y_0+1} ) \biggr)(M')^2\\
&&{} + \frac{2\delta}{\sigma^2}
(1+e^{y_0+1} )(M')^3.
\end{eqnarray*}
Hence, $|g_\Delta'(y)| \leq2M''\Delta$ for $y \in[y_0-\Delta
,y_0+\Delta]$ and $\Delta<1$. Combined with (\ref{e:odeg}), this yields
%
%
\begin{equation}\label{e:inf}
{\sup_{y \in[y_0-\Delta,y_0+\Delta]}} |g_{\Delta}''(y)| \to0\qquad
\mbox{for } \Delta\to0.
\end{equation}
For $\Delta$ sufficiently small, $y \in[y_0+\Delta,y_0+1]$, and
\begin{eqnarray*}
&&|g'_{\Delta}(y)|<m_\Delta:=\max  \Biggl\{\frac{{1}/{3}
(-{\mu}/{\sigma^2}+{1}/({1+e^{-(y_0+\Delta)}}) )}{({4\mu})/{\sigma
^2}+3+({2\delta})/{\sigma^2} (1+e^{y_0+1} )},\\
&&\hspace*{105pt}\sqrt{\frac{{1}/{3} (-{\mu}/{\sigma^2}+
{1}/({1+e^{-(y_0+\Delta)}}) )}{({2\mu})/{\sigma^2}+1+
({4\delta})/{\sigma^2} (1+e^{y_0+1} )}},\\
&&\hspace*{116pt} \sqrt[3\,]{\frac{{1}/{3} (-{\mu}/{\sigma
^2}+{1}/({1+e^{-(y_0+\Delta)}} ))}{({2\delta})/{\sigma
^2}(1+e^{y_0+1} )}} \Biggr\},
\end{eqnarray*}
(\ref{e:odeg}) and a first-order Taylor expansion imply
%
%
\begin{equation}\label{e:lbound}
g''_{\Delta}(y)>\frac{\mu}{\sigma^2}+\frac{1}{1+e^{-(y_0+\Delta
)}}>\frac
{e^{-y_0}}{2(1+e^{-y_0})^2}\Delta>0.
\end{equation}
Equation (\ref{e:inf}) yields
\[
|g_{\Delta}'(y_0+\Delta)|\leq{2\Delta\sup_{y \in[y_0-\Delta
,y_0+\Delta
]}} |g''_{\Delta}(y)|<m_\Delta
\]
for sufficiently small $\Delta$.
By (\ref{e:lbound}) we have that if $g'_{\Delta}$ does not have a zero
on $[y_0-\Delta,y_0+\Delta]$, that is, $g'(y_0+\Delta)<0$, then
$g''_{\Delta}(y)>\frac{e^{-y_0}}{2(1+e^{-y_0})^2}\Delta$ on
$[y_0+\Delta
,\min\{\overline{\beta}_{\Delta},y_0+1\}]$. Using (\ref{e:inf}),
this yields
\[
\overline\beta_\Delta-\underline{\beta}_{\Delta}<2\Delta+\frac
{{2\Delta
\sup_{y \in[y_0-\Delta,y_0+\Delta]}}|g''(y)|}{
{e^{-y_0}}/({2(1+e^{-y_0})^2})\Delta} \to0
\]
for $\Delta\to0$.
Since $|g'_{\Delta}(y)|<M'$, an application of the mean value theorem
completes the first step.
\end{Step}
\begin{Step}\label{Step2}
We now\vspace*{1pt} establish $\overline{\beta}_{\Delta} \geq y_0$ and
$g(\overline
{\beta}_{\Delta}) \to-\infty$ as $\Delta\to\infty$. To this end, let
$y^*<y_0$. Then we have $g''_{\Delta}(y)<-\frac{\mu}{\sigma
^2}+\frac
{1}{1+e^{y^*}}<0$ if $y \leq y^*$ and
\begin{eqnarray*}
&&|g'_{\Delta}(y)|<m':=\max  \Biggl\{\frac{{1}/{3} |-
{\mu}/{\sigma^2}+{1}/({1+e^{-y^*}})|}{({4\mu})/{\sigma
^2}+3+({2\delta})/{\sigma^2} (1+e^{y^*} )},\\
&&\hspace*{101.4pt}\sqrt{\frac{{1}/{3} |-{\mu}/{\sigma^2}+
{1}/({1+e^{-y^*}}) |}{({2\mu})/{\sigma^2}+1+({4\delta
})/{\sigma
^2} (1+e^{y^*} )}}, \\
&&\hspace*{118.5pt} \sqrt[3\,]{\frac{{1}/{3} |-{\mu}/{\sigma
^2}+{1}/({1+e^{-y^*}}) |}{({2\delta})/{\sigma^2}
(1+e^{y^*})}} \Biggr\}.
\end{eqnarray*}
Since $g_{\Delta}''(\underline\beta_\Delta)<0$, this implies
$g_{\Delta
}'(y)<0$ for $y \leq y^*$ as well as $|g_{\Delta}'(y)|\geq m'$ for
$y\in
[y_0-\Delta+\frac{m'}{\mu/\sigma^2-(1+e^{-y^*})^{-1}},y^* ]$. By the
first statement and since $y^*<y_0$ was chosen arbitrarily, we have
$\overline{\beta}_{\Delta} \geq y_0$. In addition, the second statement
and the mean value theorem show that $g_{\Delta}(\overline{\beta
}_{\Delta})\to-\infty$ as $\Delta\to\infty$.
\end{Step}
\begin{Step}\label{Step3}
We now establish $\overline{\beta}_{\Delta}>y_0$.
By Step \ref{Step2} it remains to show that \mbox{$\overline{\beta}_{\Delta} \neq y_0$}.
Suppose that $\overline{\beta}_{\Delta}=y_0$. Then $g'_{\Delta
}(y_0)=0=g''_{\Delta}(y_0)$ and it follows from a Taylor expansion
around $y_0$ that
\[
g''_{\Delta}(y)=\frac{2e^{-y_0}}{(1+e^{-y_0})^2}(y-y_0)+O\bigl((y-y_0)^2\bigr)<0
\]
for $y \in(y_0-\varepsilon,y_0)$ and sufficiently small $\varepsilon>0$,
hence $g'(y)>0$ for some \mbox{$y<y_0$}. By the intermediate value theorem,
there exists a zero of $g'$ on $(\underline\beta_\Delta,y_0)$, in
contradiction to the definition of $\overline{\beta}_{\Delta}$.
Therefore, we have $\overline{\beta}_{\Delta}>y_0$ as claimed.
\end{Step}
\begin{Step}\label{Step4}
Next, we prove that $(g_{\Delta},g'_{\Delta})$ converges toward
$(g_{\Delta_0},g'_{\Delta_0})$ uniformly on compacts as $\Delta\to
\Delta_0$.
To this end, we consider the solution $f^\Delta\dvtx\mathbb R _+\to
\mathbb R^3$ to the
initial value problem
\[
{d\over dy}(f^\Delta_1,f^\Delta_2,f^\Delta_3)(y)= (1,f^\Delta
_3(y),h  (f^\Delta_1(y),f^\Delta_3(y) ) )
\]
with
\begin{eqnarray*}
h(y,z)&:=&
\biggl(-\frac{2 \mu}{\sigma^2}+\frac{2}{1+e^{-y}} \biggr)+
\biggl(\frac{4\mu
}{\sigma^2}-\frac{2}{1+e^{-y}}-1-\frac{2 \delta}{\sigma
^2}(1+e^y)
\biggr)z\\
&&{}+ \biggl(-\frac{2\mu}{\sigma^2}+1+\frac{4 \delta}{\sigma
^2}(1+e^y) \biggr)z^2 -\frac{2 \delta}{\sigma^2}(1+e^y)z^3
\end{eqnarray*}
and initial values
$(f^\Delta_1,f^\Delta_2,f^\Delta_3)(0)=(y_0-\Delta,\overline{C},0)$.
The solution to this problem is
\[
(f^\Delta_1,f^\Delta_2,f^\Delta_3)(y)= \bigl(y+y_0-\Delta,g_\Delta
(y+y_0-\Delta),g'_\Delta(y+y_0-\Delta) \bigr).
\]
Note that
\begin{eqnarray*}
|g_\Delta(y)-g_{\Delta_0}(y)|&=&
 |f_2^\Delta(y-y_0+\Delta)-f_2^{\Delta_0}(y-y_0+\Delta
_0) |\\
&\le& |f_2^\Delta(y-y_0+\Delta)-f_2^{\Delta_0}(y-y_0+\Delta
)
|+M'|\Delta-\Delta_0|
\end{eqnarray*}
and similarly for $g''$. Hence, it suffices to show that $f^\Delta$
depends uniformly on compacts on its initial value $f^\Delta(0)$.

$h$ is locally Lipschitz and hence globally Lipschitz in $z$ on
$[-M',M']$ and in $y$ on compacts.
The desired uniform convergence follows now from the corollary to~\cite
{birkhoffrota62}, Theorem V.3.2.
\end{Step}
\begin{Step}\label{Step5}
In view of Steps \ref{Step1} and \ref{Step2} as well as the intermediate value theorem, it
remains to show that $g(\overline{\beta}_{\Delta})$ depends
continuously on $\Delta$. Fix $\Delta_0>0$. Since $\overline{\beta
}_{\Delta_0} > y_0$ by Step \ref{Step3}, a Taylor expansion around $\overline
{\beta}_{\Delta_0}$ yields that $g'_{\Delta_0}$ is strictly increasing
in a sufficiently small neighborhood $\mathscr{W}$ of $\overline
{\beta
}_{\Delta_0}$.
Now consider $\Delta$ sufficiently close to $\Delta_0$.
Recall that $g_\Delta'(y)$ does not vanish for $\underline\beta
_\Delta
<y\le y_0$.
By the uniform convergence from Step \ref{Step4}, the first zero $\overline
{\beta
}_{\Delta}$ of $g_\Delta'$ after $\underline\beta_\Delta$
is close to the first zero $\overline{\beta}_{\Delta_0}$ of
$g_{\Delta
_0}'$ after $\underline\beta_{\Delta_0}$.
In view of
\[
 |g_\Delta(\overline{\beta}_{\Delta})-g_{\Delta_0}(\overline
{\beta
}_{\Delta_0}) |
\le |g_\Delta(\overline{\beta}_{\Delta})-g_{\Delta
_0}(\overline
{\beta}_{\Delta}) |+
 |g_{\Delta_0}(\overline{\beta}_{\Delta})-g_{\Delta
_0}(\overline
{\beta}_{\Delta_0}) |
\]
and Step \ref{Step4}, this completes the proof.\qed
\end{Step}
\noqed
\end{pf}

We now construct the process $\beta$ as the solution to an SDE with
instantaneous reflection. The coefficients $a$ and $b$ in (\ref
{e:beta}) below are chosen in line with (\ref{e:Beta1}) and (\ref{e:sol}).
\begin{lemma}\label{l:beta}
Let $\beta_0 \in[\underline{\beta},\overline{\beta}]$ and
\[
a(y):=\frac{\sigma^2}{2} \biggl(\frac{1-e^{-y}}{1+e^{-y}}
\biggr)
\biggl(\frac{1}{1-g'(y)} \biggr)^2+\delta (1+e^y ),\qquad
b(y):=\frac
{\sigma}{1-g'(y)}
\]
for $\beta\in[\underline{\beta},\overline{\beta}]$.
Then there exists a solution to the Skorokhod SDE
\[
d\beta_t=a(\beta_t)\,dt+b(\beta_t)\,dW_t
\]
with instantaneous reflection at $\underline{\beta},\overline{\beta}$,
that is, a continuous, adapted, $[\underline{\beta},\overline{\beta
}]$-valued process $\beta$ and nondecreasing adapted processes $\Phi$,
$\Psi$ such that $\Phi$ and $\Psi$ increase only on the sets $\{
\beta
=\underline{\beta}\}$ and $\{\beta=\overline{\beta}\}$,
respectively, and
%
%
\begin{equation}\label{e:beta}
\beta_t = \beta_0 + \int_0^t a(\beta_s)\,ds + \int_0^t b(\beta_s)\,dW_s
+\Phi_t -\Psi_t
\end{equation}
holds for all $t \in\mathbb R _+$.
\end{lemma}
\begin{pf}
In view of \cite{skorohod61}, it suffices to prove that the
coefficients $a(\cdot)$ and $b(\cdot)$ are globally Lipschitz on
$[\underline{\beta},\overline{\beta}]$. By the mean value theorem
it is
enough to show that their derivatives are bounded on $(\underline
{\beta
},\overline{\beta})$. Let $y \in(\underline{\beta},\overline
{\beta})$
be fixed. Then we have
%
%
\begin{equation}\label{e:ableitung}
b'(y)=\sigma\frac{g''(y)}{(1-g'(y))^2},
\end{equation}
$g'(y)\le0$ implies $|1-g'(y)|\ge\max\{1,|g'(y)|\}$.
Moreover, $g'$ is bounded on $[\underline{\beta},\overline{\beta}]$ by
the proof of Proposition \ref{p:FBVP}.
Boundedness of $b'$ now follows from (\ref{e:odeg}) and~(\ref
{e:ableitung}). Boundedness of $a'$ is shown along the same lines.
\end{pf}

We now define $C$ and the shadow price process $\widetilde{S}$ as motivated
in Section \ref{s:heuristik}.
\begin{lemma}\label{l:shadow}
For $\beta_0 \in[\underline{\beta},\overline{\beta}]$ let $\beta
$ be
the process from Lemma \ref{l:beta}.
Then $C:=g(\beta)$ is a $[\underline{C},\overline{C}]$-valued It\^{o}
process of the form
\begin{eqnarray*}\label{e:CIto}
C_t &=& g(\beta_0)+\int_0^t \biggl(-\mu+ \frac{\sigma^2}{2}+\frac
{\sigma
^2}{2} \biggl(\frac{1-e^{-\beta_s}}{1+e^{-\beta_s}} \biggr)
\biggl(\frac
{1}{1-g'(\beta_s)} \biggr)^2 \biggr) \,ds \\
&&{} + \int_0^t\frac{\sigma g'(\beta_s)}{1-g'(\beta_s)}\,dW_s
\end{eqnarray*}
and the It\^{o} process $\widetilde{S}:= S\exp(C)$ satisfies
\[
\widetilde{S}_t = S_0 e^{C_0} \exp \biggl(\int_0^t\frac{\sigma
^2}{2}
\biggl(\frac
{1-e^{-\beta_s}}{1+e^{-\beta_s}} \biggr) \biggl(\frac{1}{1-g'(\beta
_s)} \biggr)^2\,ds+\int_0^t \frac{\sigma}{1-g'(\beta_s)}\,dW_s \biggr).
\]
\end{lemma}
\begin{pf}
$g$ can be extended to a $C^2$-function on an open set containing
$[\underline{\beta},\overline{\beta}]$, e.g., by attaching suitable
parabolas at $\underline{\beta}, \overline{\beta}$. Since $\Phi$
and $
\Psi$ are of finite variation and $g'$ vanishes on the support of the
Stieltjes measures corresponding to $\Phi$ and~$\Psi$, It\^{o}'s
formula yields
\[
dC_t =  \bigl(g'(\beta_t)a(\beta_t)+\tfrac{1}{2}g''(\beta_t)b(\beta
_t)^2 \bigr)\,dt + g'(\beta_t)b(\beta_t)\,dW_t.
\]
The claims follow by inserting the definitions of $a$ and $b$, (\ref
{e:odeg}), and the definition of $S$.
\end{pf}

Next, we show that $\widetilde{S}$ is indeed a shadow price process, i.e.,
the same portfolio/consump\-tion pair $(\varphi,c)$ is optimal with the
same expected utility both in the frictionless market with price
process $\widetilde{S}$ and in the market with price process $S$ and
proportional transaction costs $\underline\lambda,\overline\lambda
$. In
the frictionless market with price process $\widetilde{S}$, standard results
yields the optimal strategy and consumption rate.
\begin{lemma}\label{l:optimal1}
Set
%
%
\begin{equation}\label{e:raender}
\beta_0 := \cases{\underline{\beta}, &\quad if $\dfrac{\eta_S \overline
S_0}{\eta_B+\eta_S \overline S_0} <
\dfrac{1}{1+e^{-\underline{\beta}}}$,\vspace*{2pt}\cr
\overline{\beta}, &\quad if $\dfrac{\eta_S \underline S_0}{\eta_B+\eta_S
\underline S_0} >
\dfrac{1}{1+e^{-\overline{\beta}}}$.}
\end{equation}
Otherwise, let $\beta_0$ denote the $[\underline{\beta},\overline
{\beta
}]$-valued solution $y$ to
\[
\frac{\eta_SS_0e^{g(y)}}{\eta_B+\eta_SS_0e^{g(y)}}=\frac{1}{1+e^{-y}}.
\]
For processes $\beta$ and $\widetilde{S}$ as in Lemma \ref
{l:shadow}, define
\begin{eqnarray*}
\widetilde{V}_t & := & (\eta_B+\eta_S\widetilde S_0)\mathscr E
\biggl(\int_0^{\cdot
}\frac
{1}{(1+e^{-\beta_s})\widetilde S_s} \,d\widetilde{S}_s-\int_0^{\cdot
}\delta \,ds
\biggr)_t,\\
c_t & := & -\delta\widetilde{V}_t,\\
\varphi^1_t &:=& \frac{1}{1+e^{-\beta_t}} \frac{\widetilde
{V}_t}{\widetilde{S}_t},\qquad
 \varphi^0_t:=\widetilde{V}_t-\varphi^1_t\widetilde{S}_t.
\end{eqnarray*}
Then
%
%
\begin{eqnarray}\label{e:strategy}\quad
\varphi^0_t&=&\varphi^0_0-\int_0^t c_s \,ds-\int_0^t\frac{\widetilde
{V}_se^{-\beta_s}}{(1+e^{-\beta_s})^2}\,d\Phi_s+\int_0^t\frac
{\widetilde
{V}_se^{-\beta_s}}{(1+e^{-\beta_s})^2}\,d\Psi_s, \nonumber\\[-8pt]\\[-8pt]
\varphi^1_t &=&\varphi^1_0+\int_0^t\frac{\varphi^1_s e^{-\beta
_s}}{1+e^{-\beta_s}}\,d\Phi_s- \int_0^t\frac{\varphi^1_s e^{-\beta
_s}}{1+e^{-\beta_s}}\,d\Psi_s\nonumber
\end{eqnarray}
and $(\varphi,c)$ is an optimal portfolio/consumption pair with value
process $\widetilde{V}$ for initial wealth $\eta_B+\eta_S\widetilde
S_0$ in the
frictionless market with price process $\widetilde{S}$.
\end{lemma}
\begin{pf}
One easily verifies that $\beta_0$ is well defined. Moreover, we have
%
%
\begin{equation}\label{e:strategy2}
\log(\varphi^1_t)=\log(\widetilde{V}_t)- \biggl(\mu-\frac{\sigma
^2}{2}
\biggr)t-\sigma W_t-C_t-\log(1+e^{-\beta_t}).
\end{equation}
By \cite{js87}, Theorem I.4.61,
\begin{eqnarray*}
d\log(\widetilde{V}_t)
&= &  \biggl(\frac{\sigma^2}{2(1+e^{-\beta_t})^2} \biggl(\frac
{1}{1-g'(\beta
_t)} \biggr)^2-\delta \biggr)\,dt\\
&&{} + \frac{\sigma}{1+e^{-\beta_t}}
\biggl(\frac
{1}{1-g'(\beta_t)} \biggr)\,dW_t.
\end{eqnarray*}
$C$ is given in Lemma \ref{l:shadow} and for the last term in (\ref
{e:strategy2}), It\^{o}'s formula yields
\[
-d\log(1+e^{-\beta_t})= \frac{e^{-\beta_t}}{1+e^{-\beta_t}}\,d\beta
_t -
\frac{1}{2} \frac{e^{-\beta_t}}{(1+e^{-\beta_t})^2}\,d[\beta,\beta]_t.
\]
Summing up terms, we have
\[
d\log(\varphi^1_t)= \frac{e^{-\beta_t}}{1+e^{-\beta_t}}\,d\Phi_t-
\frac
{e^{-\beta_t}}{1+e^{-\beta_t}}\,d\Psi_t.
\]
Hence, $\log(\varphi^1)$ is of finite variation and another application
of It\^{o}'s formula yields the claimed representation for $\varphi^1$.
Obviously, $\widetilde V$ is the value process of $\varphi$ relative to
$\widetilde S$.
By definition, we have
%
%
\begin{equation}\label{e:sf}
d\widetilde V_t=\varphi^1_t\,d\widetilde S_t-c_t\,dt,
\end{equation}
which means that $(\varphi,c)$ is a self-financing
portfolio/consumption pair for price process $\widetilde S$.
The integral representation of $\varphi^0$ now follows from
\[
d\varphi^0_t=d(\widetilde{V}_t-\varphi^1_t\widetilde
{S}_t)=-c_t\,dt-\widetilde
{S}_t\,d\varphi^1_t,
\]
where we used integration by parts in the sense of \cite
{js87}, I.4.49b. For $t \in\mathbb R _+$ set
\[
K_t:=\int_0^t e^{-\delta s}\,ds,\qquad  \kappa_t:= e^{\delta t}
c_t,\qquad
\psi_t^0:= \varphi_t^0+\int_0^t c_s \,ds,\qquad  \psi^1_t:=\varphi^1_t.
\]
Then $(\varphi,c)$ is optimal in the sense of Definition \ref
{d:optimal} (adapted to frictionless markets where the restriction to
strategies of finite variation is dropped) if and only if $(\psi
,\kappa
)$ is optimal in the sense of \cite{kallsengoll99a}, Definition 2.2.
The differential characteristics $(\widetilde b,\widetilde c,\widetilde
F)$ of $\widetilde{S}$
are given by $\widetilde F=0$ and
\[
\widetilde b_t=\widetilde{S}_t\sigma^2\frac{1}{1+e^{-\beta_t}}
\biggl(\frac
{1}{1-g'(\beta_t)} \biggr)^2,\qquad
\widetilde c_t=\widetilde{S}^2_t\sigma^2 \biggl(\frac{1}{1-g'(\beta
_t)} \biggr)^2.
\]
Hence, \cite{kallsengoll99a}, Theorem 3.1, with $H_t=\widetilde
b_t/\widetilde
c_t$, $K_{\infty}=1/\delta$ and $K_{\infty}-K_t=\frac{1}{\delta}
e^{-\delta t}$ yields the optimality of $(\varphi,c)$.
\end{pf}

If (\ref{e:raender}) holds, then $(\varphi^0_0,\varphi^1_0)\not
=(\eta
_B,\eta_S)$.
In this case, we can and do modify the initial portfolio to
%
%
\begin{equation}\label{e:modify}
(\varphi^0_0,\varphi^1_0):=(\eta_B,\eta_S)
\end{equation}
without affecting the initial wealth, gains, or optimality.
From now on, $\varphi$ refers to this slightly changed strategy.
The case (\ref{e:raender}) happens if the initial portfolio is not
situated in the no-trade region of the transaction costs model,
which makes an initial bulk trade necessary.

(\ref{e:strategy}) implies that the optimal strategy $\varphi$ is of
finite variation and constant
until $\widetilde S$ visits the boundary $\{\underline S,\overline S\}$ the
next time.
Since sales and purchases take place at the same prices as in the
market with transaction costs $\underline\lambda$, $\overline\lambda$
and price process $S$, the portfolio/consumption pair $(\varphi,c)$ is
admissible in this market as well. Conversely, since shares can be
bought at least as cheaply and sold as least as expensively, any
admissible consumption rate in the market with price process $S$ and
transaction costs is admissible in the frictionless market with price
process $\widetilde{S}$, too. Hence, $(\varphi,c)$ is optimal in the market
with transaction costs as well. Made precise, this is stated in the
following theorem.
\begin{theorem}\label{l:optimal2}
The portfolio/consumption pair $(\varphi,c)$ defined in Lemma \ref
{l:optimal1} and (\ref{e:modify}) is also optimal in the market with
price process $S$ and proportional transaction costs $\overline\lambda
$, $\underline\lambda$. In particular, $\widetilde{S}$ is a shadow price
process in this market.
\end{theorem}
\begin{pf}
Let $((\psi^0,\psi^{\uparrow}-\psi^{\downarrow}),\kappa)$ be an
admissible portfolio/consumption pair in the market with price process
$S$ and transaction costs $\underline\lambda$, $\overline\lambda$. By
$\underline S \leq\widetilde{S} \leq\overline S$ and the self-financing
condition (\ref{e:selff2}),
\[
\widetilde{\psi}^0 := \psi^0_0+\int_0^\cdot\widetilde{S}_t\,d\psi
^{\downarrow
}_t-\int
_0^\cdot\widetilde{S}_t\,d\psi^{\uparrow}_t -\int_0^\cdot c_t\,dt \geq
\psi^0.
\]
Together with $\underline S \leq\widetilde{S} \leq\overline S$
it follows that $((\widetilde{\psi}^0,\psi^1),\kappa)$ is an
admissible\break 
portfolio/consumption pair in the frictionless market with price
process $\widetilde S$. By optimality of $(\varphi,c)$ defined in
Lemma \ref
{l:optimal1}, this implies
\[
E \biggl(\int_0^{\infty} e^{-\delta t}\log(c_t)\,dt \biggr) \geq E
\biggl(\int
_0^{\infty} e^{-\delta t}\log(\kappa_t)\,dt \biggr).
\]
Therefore, it remains to prove that $(\varphi,c)$ is admissible in the
market with price process $S$ and proportional transaction costs
$\overline\lambda$, $\underline\lambda$.
Let us begin with $\varphi$ as in Lem\-ma~\ref{l:optimal2}, that is, without
the modification from (\ref{e:modify}).
Since $\Phi$ and $\Psi$ increase only on the sets $\{\widetilde
{S}=\overline
S\}$ and $\{\widetilde{S}=\underline S\}$, respectively, the self-financing
condition for $(\varphi,c)$ and (\ref{e:strategy}) yield
\begin{eqnarray*}
\varphi^0 &=& \varphi^0_0+\int_0^\cdot\widetilde{S}_t\,d\varphi
^{\downarrow
}_t-\int_0^\cdot\widetilde{S}_t\,d\varphi^{\uparrow}_t -\int
_0^\cdot
c_t\,dt\\
&=& \varphi^0_0+\int_0^\cdot\underline S_t\,d\varphi^{\downarrow
}_t-\int
_0^\cdot\overline S_t\,d\varphi^{\uparrow}_t -\int_0^\cdot c_t\,dt.
\end{eqnarray*}
This shows that $(\varphi,c)$ is self-financing in the market with
price process $S$ and transaction costs $\overline\lambda$,
$\underline
\lambda$.
We now turn back to $\varphi$ as in (\ref{e:modify}). By definition of
$\widetilde{S}_0$, both sides of (\ref{e:selff2}) are unaffected by this
modification, at least if the initial values of $\varphi^\uparrow
,\varphi^\downarrow$ are chosen accordingly. This implies that the
slightly changed $(\varphi,c)$ is self-financing for $S,\overline
\lambda
$, $\underline\lambda$ as well.
By $\varphi^0$, $\varphi^1 \ge0$, it is also admissible. This completes
the proof.
\end{pf}

In the language of \cite{davisnorman90}, the optimal policy is
$(c,L,U)$ with
\begin{eqnarray*}
L_t&=& (\varphi^1_0-\eta_S )^+S_0+\int_0^t{\varphi
^1_sS_se^{-\beta_s}\over1+e^{-\beta_s}}\,d\Phi_s,\\
U_t&=& (\varphi^1_0-\eta_S )^-S_0+\int_0^t{\varphi
^1_sS_se^{-\beta_s}\over1+e^{-\beta_s}}\,d\Psi_s.
\end{eqnarray*}
In particular, it belongs to the slightly smaller set of admissible
controls in \cite{davisnorman90,shrevesoner94},
where the cumulative values $L,U$ of purchases and sales are supposed
to be right continuous.
Therefore, the optimal strategies in our and their setup coincide.
\begin{bem}\label{b:connectDN}
In the case of logarithmic utility, it is possible to recover the
shadow price $\widetilde{S}$ from the results of \cite{davisnorman90}.
General results on logarithmic utility maximization in frictionless
markets show that the optimal consumption rate $c$ equals the $1/
\delta
$-fold of the investor's current wealth measured in terms of the shadow
price. Hence, the consumption rate calculated in \cite{davisnorman90}
determines the shadow value process $\widetilde{V}$, which in turn
allows to
back out the shadow price $\widetilde{S}$. More precisely, the shadow price
can be constructed in a very subtle way using the results of \cite
{davisnorman90}, as was pointed out to us by the very insightful
comments of an anonymous referee: in the proof of \cite{davisnorman90},
Theorem 5.1, it is shown that the value function is of the form
%
%
\begin{equation}\label{e:vfDN}
v(x,y)=\frac{1}{\delta} \log \biggl(p \biggl(\frac{x}{y} \biggr)
\biggl(x+q \biggl(\frac{x}{y} \biggr)y \biggr) \biggr)
\end{equation}
with functions $p, q$ related through the identity
%
%
\begin{equation}\label{e:remarkable}
p'(x)=-p(x) q'(x)/\bigl(x+q(x)\bigr).
\end{equation}
Differentiating (\ref{e:vfDN}) and inserting (\ref{e:remarkable})
leads to
\[
\frac{1}{v_x(x,y)}=\delta \biggl(x+q \biggl(\frac{x}{y} \biggr)y \biggr).
\]
In view of \cite{davisnorman90}, Theorem 4.3, this shows that the
optimal consumption policy is given by $c=\delta(s_0+q(\frac
{s_0}{s_1})s_1)$. By \cite{kallsengoll99a}, Theorem 3.1, this implies
that the optimal value process w.r.t. the shadow price is given by
%
%
\begin{equation}\label{e:valueDN}
\widetilde{V}=s_0+q \biggl(\frac{s_0}{s_1} \biggr)s_1.
\end{equation}
A close look at the construction of the function $q$ in the proof of
\cite{davisnorman90}, Theorem 5.1, reveals that $q$ is increasing with
$q(\frac{s_0}{s_1})=1-\underline{\lambda}$ when $\frac{s_0}{s_1}$ hits
the lower boundary, resp., $q(\frac{s_0}{s_1})=1+\overline{\lambda}$ for
the upper boundary of the no-trade region. Therefore, it follows from
\cite{davisnorman90}, (3.1), that
\[
\widetilde{V}=(\varphi^0+\Delta s_0)+q \biggl(\frac{s_0}{s_1}
\biggr)(\varphi^1
S+\Delta s_1)=\varphi^0+\varphi^1 q \biggl(\frac{s_0}{s_1} \biggr)S
\]
for the optimal trading strategy
\[
\varphi_t^0=s_0(t-),\qquad  \varphi_t^1:=\frac{s_1(t-)}{S},
\]
corresponding to the optimal policy $(L,U)$ of \cite{davisnorman90}.
This shows that $q(\frac{s_0}{s_1})S$ coincides with the shadow price
process $\widetilde{S}$ constructed above.

However, if one wants to verify that $q(\frac{s_0}{s_1})S$ indeed is a
shadow price without using the results provided here, the ensuing
verification procedure appears to be as involved as our approach of
dealing with the utility optimization problem and the computation of
the shadow price process simultaneously. More specifically, one knows
by construction that $q(\frac{s_0}{s_1})S$ is $[(1-\underline{\lambda
})S,(1+\overline{\lambda})S]$-valued and positioned at the respective
boundary whenever the strategy $\varphi$ trades. By the proof\vspace*{1pt} of
Theorem~\ref{l:optimal2}, it therefore suffices to show that $(\varphi
,c)$ is optimal w.r.t. $\widetilde{S}=q(\frac{s_0}{s_1})S$ in order for
$q(\frac{s_0}{s_1})S$ to be a\vspace*{1pt} shadow price. In view of \cite{kallsengoll99a},
Theorem 3.1, this amounts to verifying that
%
%
\begin{equation}\label{e:optimalityDN}
\frac{\varphi^1 q(s_0/s_1)S}{s_0+q(s_0/s_1)s_1}=\frac{b}{c}
\end{equation}
for the differential semimartingale characteristics $(b,c,0)$ of the
continuous process $q(\frac{s_0}{s_1})S$. In particular, one has to
prove that the properties of the function $q$ ensure that $S$ is an It\^
{o} process and calculate its It\^{o} decomposition. The optimality
condition (\ref{e:optimalityDN}) then has to be verified using \cite
{davisnorman90}, (5.7), which leads to rather tedious computations.
Moreover, the analysis of \cite{davisnorman90} requires the technical
Condition B, which is not needed for our approach.

As a side remark, it is interesting to note that this link between
optimal policy and shadow price is only apparent for logarithmic
utility. Therefore, it is not possible to extract the shadow price from
the results of \cite{davisnorman90} for power utility functions
$u(x)=x^{1-p}/(1-p)$. Using the present approach of solving for the
optimal strategy and the shadow price simultaneously still leads to
equations for the optimal strategy and the shadow price. However, the
corresponding free boundary problem appears to be more complicated than
its counterpart in \cite{davisnorman90}. At this stage it is not
clear whether this additional complexity can be removed through
suitable transformations as in the proof of \cite{davisnorman90},
Theorem 5.1, or whether the shadow price is indeed more
difficult to obtain than the optimal policy for power utility.
\end{bem}

\section*{Acknowledgments}
We are grateful to two anonymous referees for their helpful remarks
that led, in particular, to Remark \ref{b:connectDN}.

%

%
\printaddresses

\end{document}